\documentclass{article}
\pdfoutput=1
\usepackage{graphicx} 
\usepackage{amsmath}
\usepackage{amssymb}
\usepackage[export]{adjustbox}
\usepackage{caption}
\usepackage{subcaption}

\begin{document}

\textbf{\huge Robust and efficient identification of optimal mixing perturbations
using proxy multiscale measures}
\bigskip 

C. Heffernan$^{1}$, C.P. Caulfield$^{2,1}$
\bigskip

$^{1}$Department of Applied Mathematics and Theoretical Physics, University of Cambridge,
Centre for Mathematical Sciences, Wilberforce Road, Cambridge CB3 0WA, UK\\
$^{2}$BP Institute, University of Cambridge, Madingley Road, Cambridge CB3 0EZ, UK

\textit{To appear in Philosophical Transactions of the Royal Society A}

\textit{Email correspondence ch782@cam.ac.uk}


\begin{abstract}
Understanding and optimizing passive scalar mixing in a diffusive fluid flow at
finite P\'eclet number $Pe=U h/\kappa$ (where $U$ and $h$ are characteristic velocity  and length scales, and $\kappa$ is the molecular diffusivisity of the scalar) is a fundamental problem of interest in many environmental and industrial flows. Particularly when $Pe \gg 1$,   identifying initial perturbations of given energy which optimally and thoroughly mix fluids of initally different properties can be computationally challenging. To address this challenge, we consider the identification  of initial perturbations in an idealized two-dimensional flow on a torus that extremize various measures over finite time horizons. We identify such `optimal' initial perturbations using the `direct-adjoint looping' (DAL) method, thus requiring the evolving flow to satisfy the governing equations and boundary conditions at all points in space and time. We  demonstrate that minimising  multiscale measures commonly known as `mix-norms'    over short time horizons is a computationally efficient and robust
way to identify 
initial perturbations that thoroughly mix layered scalar distributions over relatively long time horizons, provided the magnitude of the mix-norm's index is not too large. Minimisation of such  mix-norms triggers the development of coherent vortical flow structures which effectively mix, with the particular properties of these flow structures
depending on $Pe$ and also the time horizon of interest.
\end{abstract}

\section{Introduction}
Understanding how fluids mix together is of central importance in the modelling of the climate, the environment, and myriad industrial processes. Mixing of a passive scalar is composed of an initial stirring action and then diffusion \cite{1} to homogenize the concentration field. At finite P\'eclet number, $Pe$ (loosely the ratio of advective to diffusive transport, more precisely defined below), diffusion is stronger in areas of high shear caused by the stirring in a mechanism known as `Taylor dispersion' \cite{2,3}. However, there is as yet no  unified mathematical theory and thus quantifying and describing the phenomenon of mixing is still of research interest. 

A first step in describing mixing is how to quantify the `mixedness' of a passive scalar. Classically, the $L^2$ norm has been used as when applied to a mean-zero field the variance is essentially a measure of homogenization \cite{4}.
%
%
 Mathematically, this choice of norm fails in the absence of diffusion and so a different measure must be sought \cite{5}. This issue motivated the development of the so-called \textit{mix-norm}, 
 as a way to quantify mixing using a Sobolev norm of index $-s=-\frac{1}{2}$ \cite{6}. This work was extended to show that any negative index Sobolev norm (i.e. for a variety of values of  the index $s$) is consistent with the rigorous ergodic theory of mixing \cite{7}.
 The mix-norm is a Sobolev norm of negative index, and here we define it  for a passive zero-mean scalar $\theta$ on a 2D torus $\Omega$ as
 \begin{align}
     ||\theta||^2_{H^{-s}(\Omega)} &= \sum_{\mathbf{k}\neq {\mathbf{0}}} |\mathbf{k} |^{-2s} |\hat{\theta}_{\mathbf{k}}|^{2},
 \end{align}
 where $s$ is the \textit{index}, $\mathbf{k}$ is the wave vector and $\hat{\theta}_{\mathbf{k}}$ are the  Fourier coefficients of $\theta$. Definition (1.1) will be the mix-norm we refer to for the remainder of this paper with the variance (for zero-mean scalars) corresponding to the special case  $s=0$\cite{5}.

Quantifying mixing is also clearly necessary 
in circumstances where it is of interest either to enhance or suppress mixing.
It has been conjectured that an appropriate way to  enhance mixing is through maximization of the time-averaged energy growth using a cost functional approach \cite{8,9}. While this does indeed lead to a well-mixed result, the problem is actually not designed to maximize mixing directly and it is natural to ask whether a more direct approach can lead to a more `efficient' well-mixed solution with less energy injection. In order to examine this problem, an optimization method based on the fully nonlinear Navier-Stokes equations is desirable. Such a method involving `adjoint'  Lagrange multipliers (that impose the governing equations) has been developed to consider a variety of optimization problems arising in fluid dynamics \cite{10,11}, and is often referred to as the `direct-adjoint-looping' (DAL) method.

In particular, the DAL method \cite{12} has been used to study mixing problems in a variety of flows \cite{13,14,15}. These studies have been based on a minimization of the mix-norm (typically with $s=1$) and compared to the results for variance minimization as well as energy maximization. It was observed that inferior mixing of the passive scalar occurs in the case of energy growth maximization for perturbations of the same initial energy. It was also shown that such mix-norm minimization appears to act well as a proxy for the variance-optimized strategy at finite target time, which is the natural measure for the $Pe<\infty$ case. Specifically, perturbations that minimized the mix-norm (with index $s=1$) over relatively
short target times led to  time-variation of the variance (and hence the mixing properties of the flow) typically very similar to the perturbations 
that minimized the variance over relatively long target times. This apparent property  has several attractions, in particular in that
the required computational demands using the DAL method are significantly smaller
and more robust for the short-target-time {mix-norm} calculations
compared to long-target time variance calculations, especially for flows at higher $Pe$.

In this paper, we are interested in answering three questions which have naturally arisen from these previous studies. Firstly, \textit{are there a range of indices $s$ which make the mix-norm a good proxy (in the above computationally efficient and robust sense) for variance-based strategies?} We find that there are indeed indices which work well in this respect while others are not suitable, as they lead to sub-optimal flow behaviour,
which we refer to as `{demixing}', and is physically associated with long-lived vortical coherence in the time-evolving flow. Secondly, \textit{how does the choice of index qualitatively change the mixing dynamics of the flow?} It is observed that changing the index can deprecate or enhance small-scale structures which can enhance or deprecate the mixing, depending on the particular choice of index $s$. Thirdly, \textit{what comparisons, if any, can be made between the variation of the index at relatively low and high $Pe$?} As one might expect, the higher P\'eclet number flows tend to favour small-scale structures, due to the lesser initial influence of diffusion. This actually makes the phenomenon of {demixing} more clearly apparent at higher $Pe$, and the observation of deprecating certain scales leading to inferior mixing holds even more clearly than in the lower P\'eclet number case. 

The rest of this paper is structured as follows. In section 2, we state the problem and methodology following the work of \cite{16}. In section 3 we present the results of our work. Specifically, in section 3 (i) we study the use of  mix-norm with various indices as  proxies for variance minimization. In particular, we show that a flow field which minimizes a mix-norm for relatively small target time can lead to similar and in some cases eventually superior mixing to those fields that minimize variance at larger times. In section 3 (ii), we show the results of how changing the index $s$ and the target time corresponds to a qualitative change in the mixing paradigm at a relatively low choice of $Pe$. We discuss analogous results for flows at higher $Pe$ in section 3 (iii), comparing the two cases. Finally, in section 4, we present brief conclusions and suggest further avenues for future study.


\section{Methods}
There are three control parameters which determine the optimal initial flow field in this paper. These are the mix-norm index $s$, the target time  $T$ of the optimization problem, and the P\'eclet number $Pe$ (defined below). We keep the Schmidt number $Sc=\nu/\kappa=1$ 
(where $\nu$ is the kinematic viscosity and $\kappa$ is the scalar diffusivity) fixed, so that the flow Reynolds number $Re=Pe$,  and also fix the initial perturbation energy density. Solutions of the optimization algorithm corresponding to these parameters will be denoted by $OA(s, T, Pe)$. We choose to study 
problems with  $Pe=50$ and $500$, $s= 0.5,1,2,5$ for the index of the mix-norm, and $T = 0.5,1,2, 5$. We also consider for comparison the behaviour of solutions for the variance with target time $T=5$ for the two choices of $Pe$, i.e. $OA(0,5,50)$ and $OA(0,5,500)$.

We use the nonlinear DAL method \cite{12} to compute the initial velocity field which will minimize the value of the mix-norm (for various indices $s$) and variance (i.e. for $s=0$) at a given target time $T$. The flow takes place in a 2D torus of length $2 \pi$ with $x$ and $y$ denoting the horizontal and vertical directions respectively. The velocity field $\mathbf{u}=(u,v)$ and pressure $p$ are governed by the incompressible Navier-Stokes equations and the passive scalar field $\theta$ is governed by a conventional advection-diffusion equation. Therefore, the non-dimensionalized equations governing the evolution of these variables are:

\begin{align}
    \frac{\partial \mathbf{u}}{\partial t} + \mathbf{u}\cdot \nabla \mathbf{u} &= -\nabla p + Re^{-1} \nabla^2 \mathbf{u}, \\
    \nabla \cdot \mathbf{u} &= 0, \\
    \frac{\partial \theta }{\partial t} + \mathbf{u} \cdot {\nabla}\theta &= Pe^{-1} \nabla^2 \theta,
\end{align}
where $Pe$ and $Re$ denote the P\'eclet number and Reynolds number respectively and are defined by 
\begin{equation}
    Pe = \frac{U h}{\kappa},\
    Re = \frac{U h}{\nu},
\end{equation}
where $U, h$ are the characteristic velocity and length scales and $\nu, \kappa$ are the kinematic viscosity and scalar diffusivity respectively.
As already noted, since we require $Sc=\nu/\kappa=1$, $Re=Pe$. 

As the scalar field is passive,   we may solve equations (2.1)-(2.2) and (2.3) separately. As an initial scalar distribution, we choose \begin{align}
   \theta(\mathbf{x},0) := \theta_0(\mathbf{x}) &= \tanh\left(6\left(x - \frac{\pi}{2}\right)\right) -  \tanh\left(6\left(x - \frac{3\pi}{2}\right)\right) - 1.
\end{align}
This corresponds to a smooth zero-mean scalar distribution with a vertical stripe, centred at $x=\pi$ of width  $\pi$ of positive $\theta \simeq 1$, bordered by stripes of negative $\theta \simeq -1$.  We also require an initial condition for the velocity field $\mathbf{u_0} = \mathbf{u}(\mathbf{x}, 0)$ in order to solve the system. For the very first loop of the DAL method, we set $\mathbf{u_0}$  to be random noise. After each iteration of the loop, we update  $\mathbf{u_0}$ to give us our initial condition to evolve the system (2.1)-(2.3).

In order to identify the initial perturbation which optimizes the mixing of the initially striped fluid, we seek to minimize an objective functional subject to a  constraint on the kinetic energy of the initial perturbation: 
\begin{align}
    ||\mathbf{u_0}||^2_{L^2(\Omega)} &= 2 e_0 \mu(\Omega), 
\end{align}
where $e_0=0.03$ is the perturbation energy density and $\mu$ denotes the `volume' (i.e. the area) of the flow domain $\Omega$. This perturbation energy density is chosen to be sufficiently large to allow for the identification 
of non-trivial initial flow structures, and yet sufficiently small so that there is still the possibility to distinguish the mixing efficacy of different initial perturbation structure.

We define the objective functional as 
\begin{align}
    \mathcal{J}(\theta (T)) = \frac{1}{2}||\theta(\mathbf{x}, T)||_{H^{-s}(\Omega)}^2 ,
\end{align}
i.e. (half) the value of the Sobolev norm of (negative) index $-s$ (which we refer to as the mix-norm of index $s$) at the target time $T$.
We may then define the constrained optimization problem of interest as \begin{align}\text{argmin } \mathcal{J}(\theta (T)) \text{ subject to } ||\mathbf{u_0}||^2_{L^2(\Omega)} = 2 e_0 \mu(\Omega), 
\end{align}
where $\{\mathbf{u}, \theta\}$ solve the system (2.1)-(2.3). The initial condition $\mathbf{u}_0$  does not appear explicitly in the objective functional, but nevertheless it affects $\mathcal{J}$ through the evolution of the flow variables which are constrained by the system (2.1)-(2.3). These constraints must be imposed of course.  This is done by the use of Lagrange multipliers, the spatially and temporally evolving so-called \textit{adjoint} variables denoted by $\{\mathbf{u}^{\dagger}, p^{\dagger}, \theta^{\dagger}\} = \{\mathbf{v}, q, \eta\}$. This is explained in detail in (for example) \cite{16}  and we follow their approach here. We may define a Lagrangian as

\begin{align}
    \mathcal{L} &= \mathcal{J}(\theta (T)) - \Sigma_{I \in \{NS, AD, C, IC\}}\quad \mathcal{J}_I,
\end{align}
where
\begin{align}
    \mathcal{J_{NS}} &= \int_0^T\int_{\Omega} \mathbf{v} \cdot \left( \frac{\partial \mathbf{u}}{\partial t} + \mathbf{u}\cdot \nabla \mathbf{u} +\nabla p - Re^{-1} \nabla^2 \mathbf{u}\right), \\
    \mathcal{J_{AD}} &= \int_0^T\int_{\Omega} \eta \left(\frac{\partial \theta }{\partial t} + \mathbf{u} \cdot {\nabla} \theta - Pe^{-1} \nabla^2 \theta \right), \\
    \mathcal{J_{C}} &= \int_0^T\int_{\Omega} q \nabla \cdot \mathbf{u}, \\
    \mathcal{J_{IC}} &= \int_{\Omega} \mathbf{v_0} \cdot (\mathbf{u}(\mathbf{x}, 0) - \mathbf{u_0}).
\end{align}
Variation with respect to adjoint variables yields equations (2.1)-(2.3). Similarly, variation with respect to the direct variables $\{\mathbf{u}, p, \theta\}$ results in the so-called \textit{adjoint Navier-Stokes} equations
\begin{align}
    \frac{\partial \mathbf{v}}{\partial t} + \mathbf{u}\cdot \nabla \mathbf{v} &= -\nabla q - Re^{-1} \nabla^2 {\mathbf{v}} + \eta\nabla \theta, \\
    \nabla \cdot \mathbf{v} &= 0, \\
    \frac{\partial \eta }{\partial t} + \mathbf{u} \cdot \nabla \eta &= - Pe^{-1} \nabla^2 \eta.
\end{align}
At $t = 0, T$ we also produce the following terminal and initial conditions
\begin{align}
    \mathbf{v}(\mathbf{x}, T) &= 0, \\
    \eta(\mathbf{x}, T) &= \sum_{\mathbf{k}\neq\mathbf{0}} |\mathbf{k}|^{-2s}Re\{ \hat{\theta}_{\mathbf{k}}(T)\exp{(i \mathbf{k}\cdot\mathbf{x})}\} , \\
    \mathbf{v_0} &= \mathbf{v}(\mathbf{x}, 0), \\
    \nabla_{\mathbf{u_0}} \mathcal{L} &= \mathbf{v_0},
\end{align}
where $\hat{\theta}_{\mathbf{k}}$ are the Fourier coefficients of $\theta$ and the $Re\{\cdot\}$ denotes real part. Due to the negative diffusion terms $-Re^{-1}\nabla^2\mathbf{v}$ and $-Pe^{-1}\nabla^2\eta$, equations {(2.14)-(2.16)} must be integrated backwards in time to avoid numerical instability. These equations are then integrated backwards from $t = T$ to $t = 0$,  thus forming 
a `direct-adjoint loop'.  Using a numerical technique from \cite{17} and with $\mathbf{u_n}$ and $\mathbf{v_n}$ known as the direct and adjoint velocities at $t = 0$ after $n$ loops of this direct-adjoint-looping (DAL) method, the updated guess $\mathbf{u_{n+1}}$ can be calculated by $$\mathbf{u_0}^{n+1} = \cos(\phi) \mathbf{u_0}^n + \sin(\phi) \mathbf{w}^n,$$ 
where $\mathbf{w}$ denotes the scaled (by the energy constraint) adjoint veclocity projected onto the hypersurface tangential to the energy hypersphere at $\mathbf{u_0}^n$, as described in detail in \cite{18}. The angle of rotation $\phi$ is calculated by using a backtracking line search \cite{19}. This looping procedure is repeated until convergence has been reached as measured by the normalized residual $r$, defined by 
$$r = \frac{||\nabla_{\mathbf{u_0}}\mathcal{L} ^{\perp}||_{L^2(\Omega)}^2}{||\nabla_{\mathbf{u_0}}\mathcal{L} ||_{L^2(\Omega)}^2},$$
where the symbol $\perp$ denotes projection onto the hyperplane tangential to the energy hypersurface. Since the energy is fixed, a small residual ($r \sim O(10^{-3})$) implies the gradient can only change by varying its magnitude which is not permissible due to the (explicitly imposed) energy constraint.

The direct and adjoint equations are solved with a $4^{th}$-order mixed Crank-Nicholson Runge-Kutta scheme with incompressibility enforced through a fractional step method \cite{13,20}. Simulations for $Pe=50$ and $Pe=500$ were performed with $N=128$ and $N=256$ grid points in both directions respectively. For the plots, the mix-norm and variance are scaled by the evolution of the purely diffusive passive scalar defined as 
\begin{align}
    M_s(t) &= \frac{||\theta(\mathbf{x},t)||^2_{H^{-s}(\Omega)}}{||\theta_d(\mathbf{x},t)||^2_{H^{-s}(\Omega)}}, \label{eq:msdef}\\
    V(t) &= \frac{||\theta(\mathbf{x},t)||^2_{L^2(\Omega)}}{||\theta_d(\mathbf{x},t)||^2_{L^2(\Omega)}},\label{eq:vdef}
\end{align}
where $\theta_d$ is the solution of equation (2.3) but with the advective term dropped, and the dependence of the (scaled) mix-norm on the index $s$ is labelled by the subscript.


\section{Results}

We now analyse the data obtained for various control parameter combinations. Mixing is measured by (scaled) variance decay as defined in equation (\ref{eq:vdef}), corresponding to the homogenization of the passive scalar.
\begin{figure}[h!]
\centering
\includegraphics[width=0.9\textwidth]{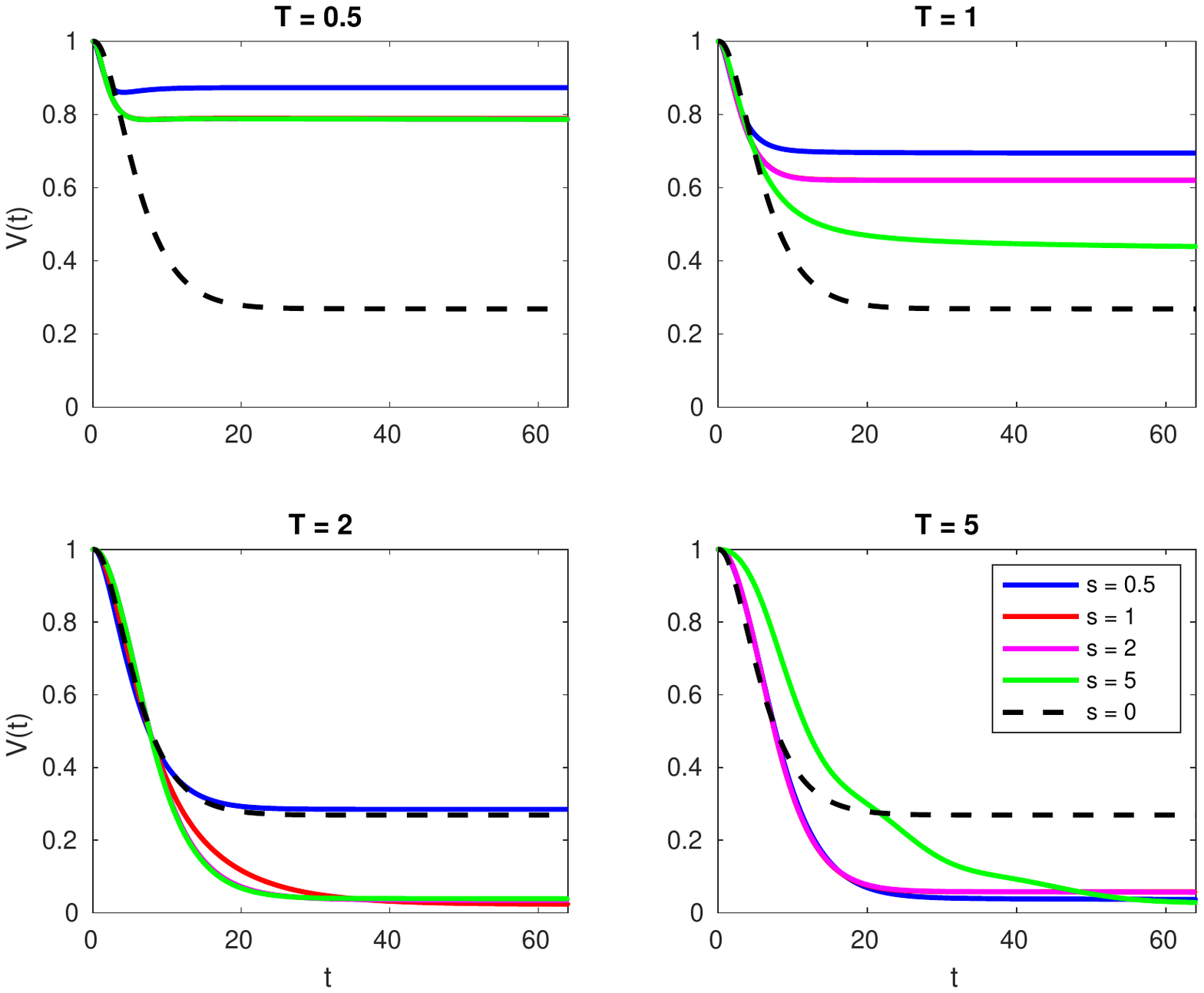}
\caption{Evolution with time of scaled variance $V(t)$, as defined in (\ref{eq:vdef})  for flows initially seeded with perturbations that minimize mix-norms (for a variety of the values of index $s$) over different target times $T$ for flows with $Re=Pe=50$. For comparison, the dotted black line shows the evolution for the perturbation that minimizes the variance for $T=5$ in a flow with $Re=Pe=50$.}\label{fig:var50}
\end{figure}
\begin{figure}[h!]
\centering
\includegraphics[width=0.9\textwidth]{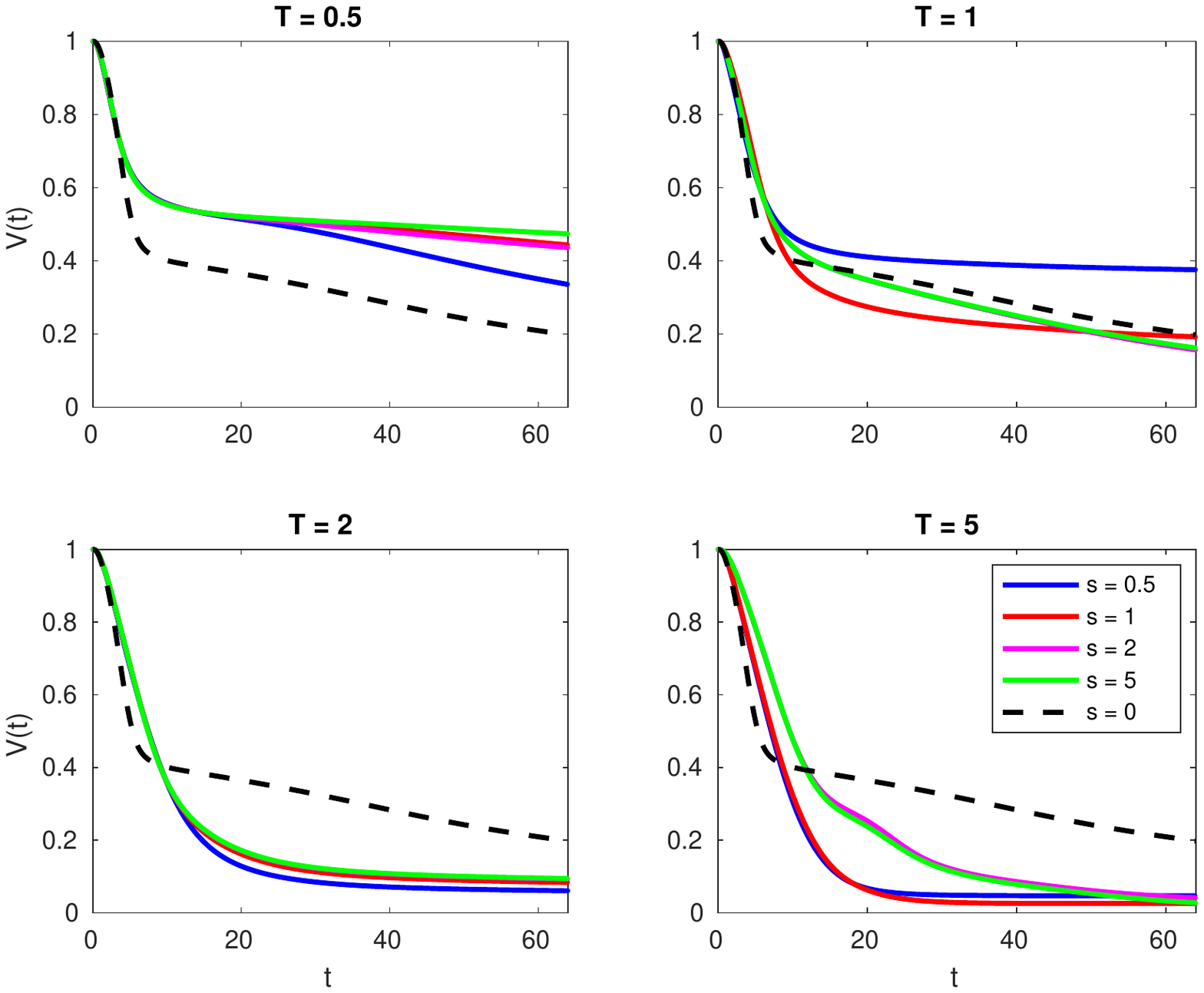}
\caption{Evolution with time of scaled variance $V(t)$, as defined in (\ref{eq:vdef}) for flows initially seeded with perturbations that minimize mix-norms (for a variety of the values of index $s$) over different target times $T$ for flows with $Re=Pe=500$. For comparison, the dotted black line shows the evolution for the perturbation that minimizes the variance for $T=5$ in a flow with $Re=Pe=500$.}\label{fig:var500}
\end{figure}


\subsubsection{Mix-Norm as a Proxy}
The mix-norm was introduced to resolve the issue with variance as a mixing measure in non-diffusive systems\cite{5}. In previous studies, the fields calculated via mix-norm minimization over short target times approach the long-time behaviour of  variance-minimizing flows with  significantly larger target times. This is an attractive feature of using mix-norms as   objective functionals as they can produce very good (and robust) approximations to exact variance based strategies over longer target times but at significantly cheaper computational cost. There are (apparently) two reasons for this cheaper computational cost. First, integrating around loops  with shorter target times clearly is cheaper than integrating around longer time loops. Secondly, and somewhat more subtly, it appears that mix-norm iterations converge more rapidly towards the required optimal solution, apparently due to more efficient identification of 
appropriate flow structures which mix well. This also contributes to the `robustness' of the method, in that the initial perturbations identified for shorter target time problems are still set  `on the right path' through time and continue to mix (through rapid variance reduction) for times significantly longer than the imposed target time. 

These characteristics are presumably related to the fundamental attractive property of multiscale measures such as mix-norms, in that large scales in the scalar distribution will correspond to large values of the mix-norm. Therefore, searching for flows that minimize mix-norms will tend to deprecate large scales within the flow, thus encouraging a cascade to smaller scales more conducive to homogenization and mixing. 
In this section, we investigate the mix-norm as a mixing proxy (in the sense described above) for 
various values of the index $s$.

Figures \ref{fig:var50} and \ref{fig:var500}
show this comparison at $Pe=50$ and $Pe=500$ respectively. As can be seen from early times, the mix-norm optimal perturbations (for various $s$) follow a very similar path to the variance-optimized perturbations with $T=5$,
consistently with previous studies \cite{13,14,15}. Interestingly, for $T \geq 2$, and a range of $s$, the mix-norm-optimized perturbations significantly
outperform the variance-optimized perturbations for times appreciably longer than its target time of $T=5$, demonstrating the (valuable) robustness of using such mix-norms
in the DAL method. 

In the case of $Pe=50$, combining a high index with sufficiently long target time gives the best proxy. This is apparent for $s=1,2$ with $T = 2$, for example, as the initial variance decay in these cases is very similar to the variance-optimizing perturbation $OA(0,5,50)$ but continues to decay significantly beyond $t = 5$, implying a higher quality mixture. Clearly, then, this near identical initial behaviour followed by further mixing gives further evidence that the mix-norm is an excellent proxy for the variance measure and motivates its use in optimization problems. However, one must be cautious with the index and target time choice, as certain combinations
perform better than others.


The mix-norm also appears to be a a good proxy  at larger $Pe$. As shown in figure \ref{fig:var500}, for flows with $Pe=500$, we can see that the optimal perturbation $OA(0,5,500)$ has the steepest decay for early times, unlike in the $Pe=50$ case where all fields had a similar decay. In fact, we observe that the variance-optimal field continues to decay up until at least the time $t=64$. This is in contrast to what was observed at $Pe=50$ where the plot approaches a constant value, indicating purely diffusive mixing, with no advection-enhanced homogenization. This is due to the persistence of  small scale structures in flows at higher $Pe$ which continue mixing after the chosen target time. 

For later times $t \gtrsim 5$, there are also choices of $s$ which yield better mixing properties than the variance-based optimal perturbations. Mix-norms  with any of the choices of index with a large enough target time appear to out-perform the mixing properties  of the  flow associated with the variance-optimal perturbation $OA(0,5,500)$. Interestingly, in some cases particuarly with smaller target times, such as $OA(1,1,500)$, these superior mixing properties are only temporary, eventually being outperformed by the mixing properties associated with the vaiance-optimal perturbation $OA(0,5,500)$. Furthermore, in the cases of larger indices for $T=5$, i.e.  the optimal perturbations $OA(2, 5, 500)$ and $OA(5,5,500)$ there are clear \textit{bumps} in the scaled variance decay for $15\lesssim t\lesssim 25$, which   will be discussed in the next sections. Using any of the indices with the intermediate target time $T = 2$ proves to be a good proxy for the variance-based strategy and, as in the lower $Pe$ case, can actually lead to superior mixing properties at later times. 


\subsubsection{Mix-norm evolution for $Pe=50$}

\begin{figure}[h!]
\centering
\includegraphics[width=0.9\textwidth]{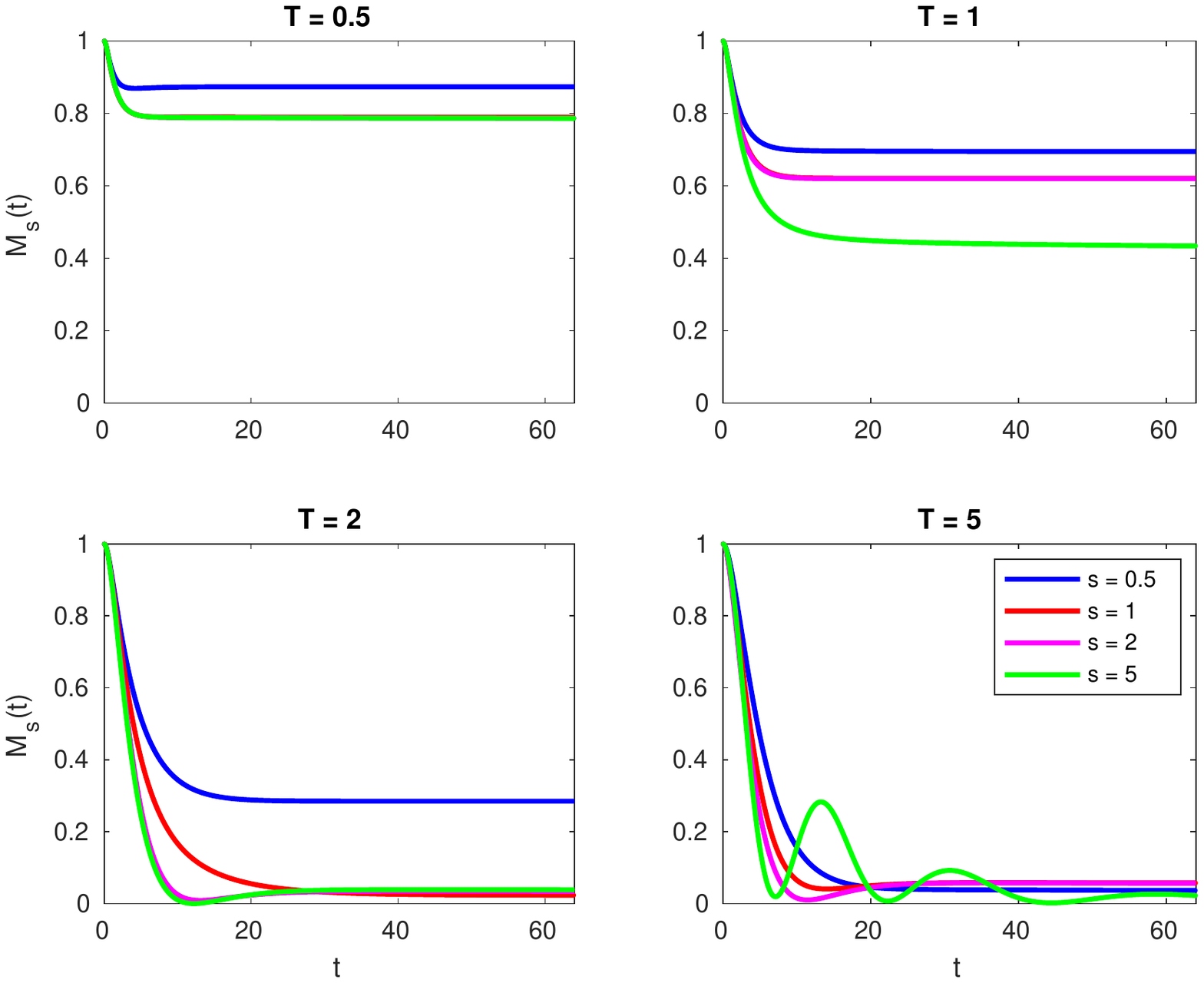}
\caption{Evolution with time of scaled mix-norm $M_s(t)$, as defined in (\ref{eq:msdef}) for flows initially seeded with perturbations that minimize mix-norms (for a variety of the values of index $s$) over different target times $T$ for flows with $Re=Pe=50$.  }\label{fig:mix50}
\end{figure}
We now analyze in more detail how the time evolution of the mix-norm differs qualitatively with index  for flows with the  lower P\'eclet number $Pe=50$, as shown in figure \ref{fig:mix50}. We observe that increasing target time $T$ leads to smaller values of the mix-norm across all values of $s$. 
However, larger values of $s$ also appear to produce a qualitative change in the dynamics. For example, despite approaching pure diffusion as $t\rightarrow\infty$ , the time evolution of $M_5(t)$ for the optimal inital perturbation $OA(5,5,50)$ exhibits \textit{demixing}, in the specific sense that $M_s(t)$ does not monotonically decay at intermediate times as seen in figure \ref{fig:mix50}.  This is an undesirable quality as a solution is only ergodic mixing if the mix-norm decays with time, and appears to be associated
with the (relatively) poor decrease in the variance
for this perturbation as can be seen in the fourth panel of figure \ref{fig:var50}.

Studying the plots of time variation of scaled mix-norm and variance proves that the choice of control parameters can produce qualitatively different results for mixing fluids. It is therefore natural to ask exactly how these choices, particularly for the index $s$, determine the structure of the initial condition $\mathbf{u_0}$ and thus at what scales mixing occurs. A deeper understanding of this question can distinguish between desirable and undesirable structures to homogenize a particular passive scalar distribution. A natural way to consider the flow dynamics is to calculate the vorticity, $\omega$, defined for such a two-dimensional flow as
\begin{align}\label{1.1}
\begin{split}
\omega = \frac{\partial v}{\partial x} - \frac{\partial u}{\partial y}.
\end{split}
\end{align}
\begin{figure}[!h]
    \centering
    \includegraphics[width=0.9\textwidth]{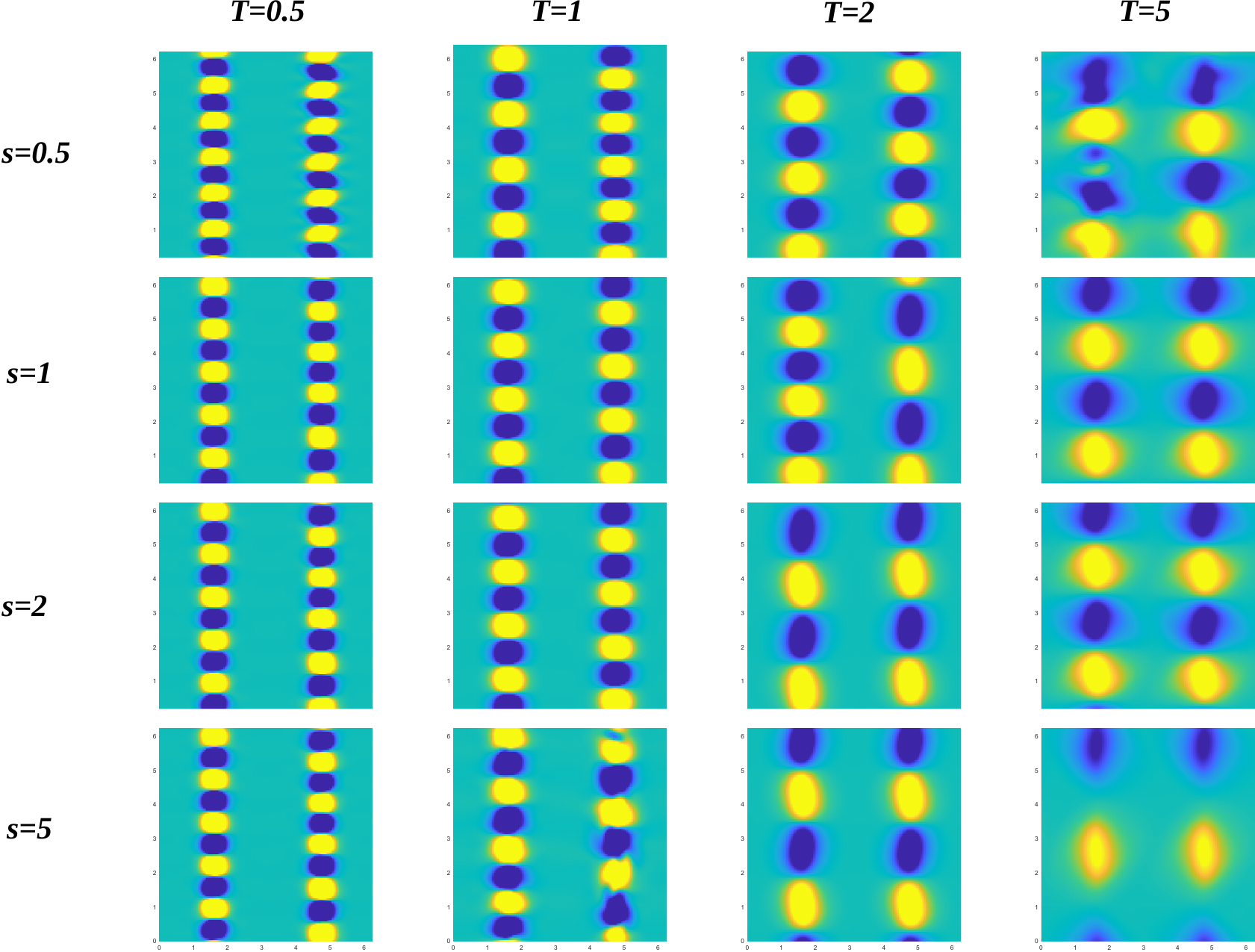}
    \hspace*{1.2cm}
    \includegraphics[width=0.6\textwidth]{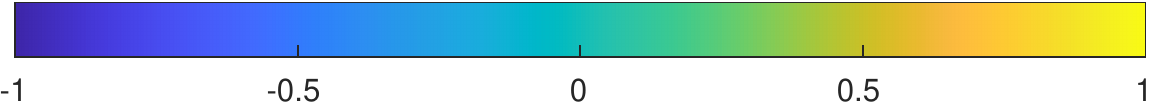}
\caption{Vorticity distribution of the optimal 
perturbations identified for mix-norm minimization for flows with $Pe=50$ with various combinations of index $s$ (rows) and target time $T$ (columns). }\label{fig:vort50}
\end{figure}
The vorticity for the various initial perturbations calculated for flows with $Pe=50$ are shown in figure \ref{fig:vort50}. These plots show two different types of initial structure. The first type, associated with optimizations for low index $s$ and target time $T$, has a large number of small scale alternating sign vortices arranged along the interfaces of the passive scalar. The second type, associated with optimizations for  high index $s$ and $T$, has a significantly smaller number of larger scale alternating sign vortices along the interface. This trend also appears to hold when varying just one of $s$ and $T$ and keeping the other parameter fixed. Comparing figure \ref{fig:vort50} with the plots in figure \ref{fig:var50}, it is clear that more vortices yield the optimal result over shorter times,  but actually lead to less thorough mixing at long times. Increasing the initial size and reducing the number of the vortices leads to  lower variance at a later time. However, there is a `sweet spot', as if there are too few vortices then the demixing phenomenon occurs and the homogenization process actually slows and leads to weaker mixing overall. 

\begin{figure}[!h]
    \centering
    \includegraphics[width=0.9\textwidth]{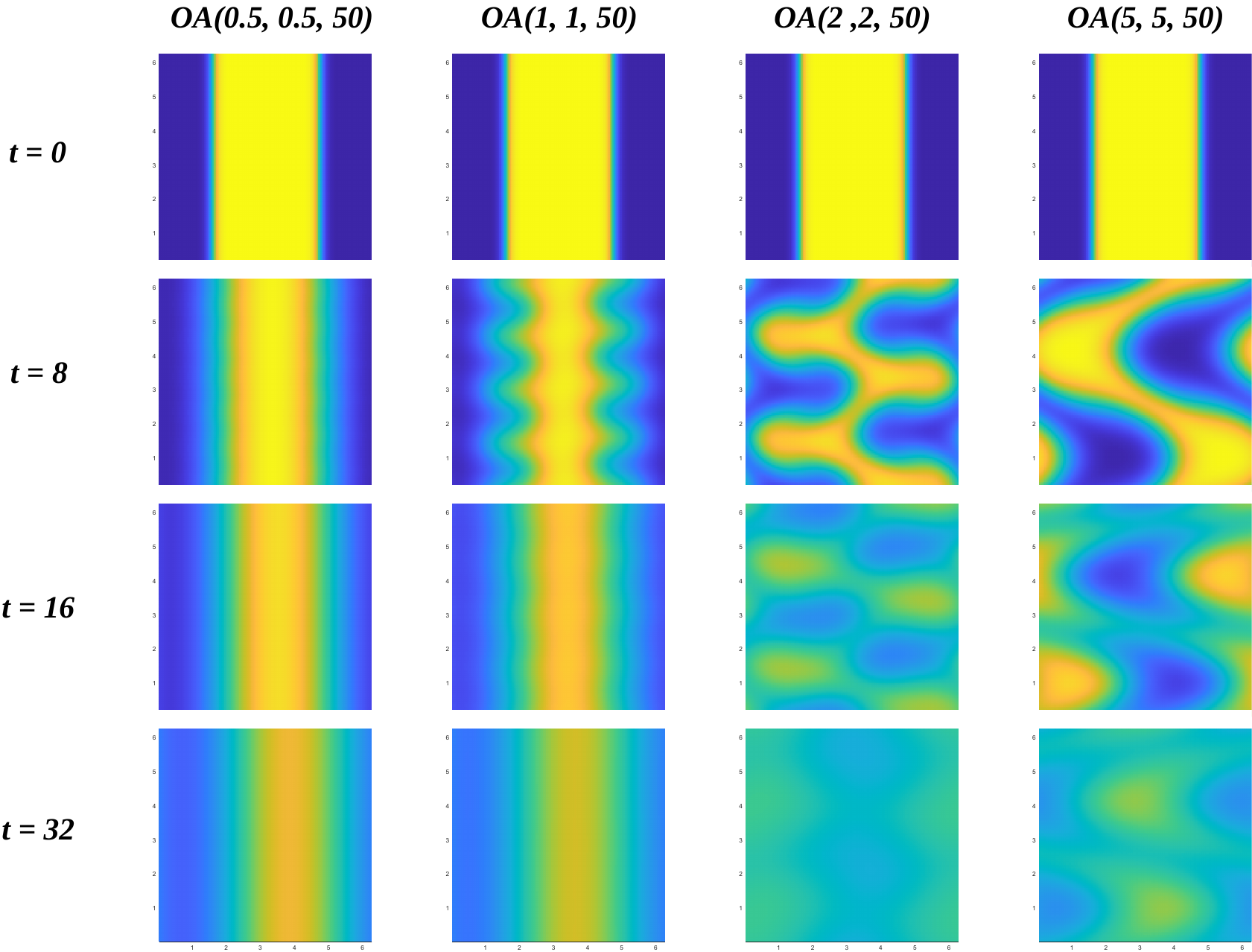}
    \hspace*{1.2cm}
    \includegraphics[width=0.6\textwidth]{colorbar-crop.pdf}
\caption{Snapshots at various times (rows) of  the evolution of the passive scalar fields mixed by   optimal initial perturbations $OA(s,T, 50)$ for various choices of $s=T$ (columns). }\label{fig:diag50}
\end{figure}
To show how these initial distributions actually  affect the time-dependent mixing dynamics, we plot in figure \ref{fig:diag50} the evolution of the passive scalar at various snapshots during the flow (with animations available as supplementary materials). We consider the evolution of the flows associated with the initial conditions shown along the diagonal of figure \ref{fig:vort50}. For the flow associated with the optimal perturbation $OA(0.5,0.5,50)$, shown in the first column of figure \ref{fig:diag50}, we see the large number of small scale vortices rapidly expend the available kinetic energy to distort  the two interfaces leading to the dominance of diffusion for the rest of the evolution. Significantly, the vortices are too small to disrupt completely the initial vertically-striped structure, and the vertical striping survives to later times. This dynamical evolution is largely similar to the
evolution of the flow associated with $OA(1,1,50)$, as shown in the second column. However, for this flow, 
the vortices in this case are fewer and larger which leads to more disruption of the interfaces between the regions of high and low concentration and thus a somewhat better mixing outcome at later times as shown in figure \ref{fig:var50}. In both of these cases the kinetic energy is still used up too quickly to disrupt completely the central stripe and so diffusion becomes the dominant factor in the mixing process early in the evolution. 

The behaviour of the other two flows is qualitatively different.  The vortices associated with the initial optimal perturbation $OA(2,2,50)$ clearly act on a larger scale, and in particular the dynamics generated manage to fold and stretch the interfaces to break the central stripe with diffusion dominating at later times leading to a close to well-mixed scalar. The flow induced by the initial perturbation $OA(5,5,50)$ (shown in the right-most column) is similar,  but in this case the lower number of initial vortices do not disrupt the central stripe as quickly. The passive scalar is homogenized quite well but the final panel shows that the larger vortices have actually led to sustained patchiness (and hence poorer mixing) for the flow
associated with the largest index $s=5$. This is a manifestation of the demixing phenomenon mentioned above, in particular in that the originally negative values of the scalar, initially associated with the edges of the flow domain (and coloured blue) have been advected in the central region, without being thoroughly mixed with the positive scalar (coloured yellow) which conversely has been advected from the centre to the edges of the flow domain. 

\begin{figure}[!h]
    \centering
    \includegraphics[width=0.9\textwidth]{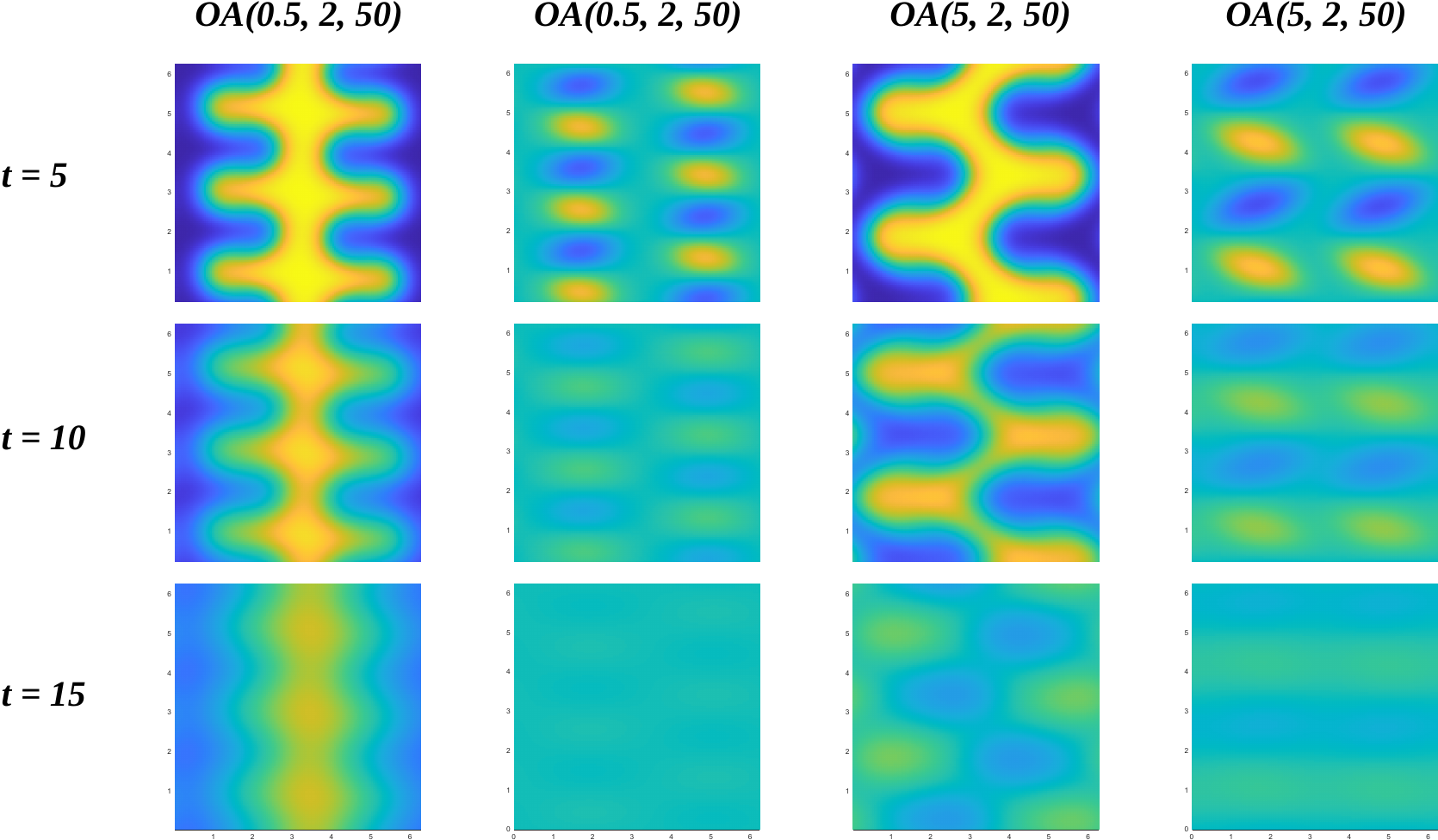}
    \hspace*{1.2cm}
    \includegraphics[width=0.6\textwidth]{colorbar-crop.pdf}
\caption{Snapshots of the evolution of the passive scalar (left) and vorticity (right) for the flows associated with the optimal perturbations $OA(0.5,2,50)$ on the first two columns and $OA(5,2,50)$ on the second two columns. Snapshot times are $t=5,10,15$.}\label{fig:t250}
\end{figure}
We also note that there are some qualitative differences in the symmetry of the optimal initial perturbations identified by the DAL method. A particular clear example is shown in the $T=2$ column of figure \ref{fig:vort50}, where the optimal
perturbation $OA(0.5,2,50)$ is anti-symmetric about the vertical mid-line while the optimal perturbation $OA(5,2,50)$ is symmetric.  It is worth asking whether or not these symmetry properties have an important role in the mixing of the passive scalar. To investigate this, we plot in figure \ref{fig:t250} the evolution of the passive scalar (left columns) and the vorticity (right columns) for the flows associated with the $s=0.5$ (anti-symmetric) and $s=5$ (symmetric) indices for $T=2$ (with animations available as supplementary materials). As can be seen from these plots, the symmetrically-aligned vortices mix the passive scalar essentially anti-symmetrically, while the anti-symmetric vortices mix the passive scalar essentially symmetically.   Furthermore, the symmetrically-aligned vortices break the central stripe somewhat more easily as it results in more stretching and folding and hence both filamentation and regions of high scalar gradient, naturally  conducive to enhancement of mixing at later times. Conversely,  the  initially anti-symmetrically aligned vortices lead to counter-rotating vortices approaching one another along the same horizontal level, making the folding more difficult, and so the mixing less thorough.




\subsubsection{High P\'eclet Number}

\begin{figure}[h!]
\centering
\includegraphics[width=0.9\textwidth]{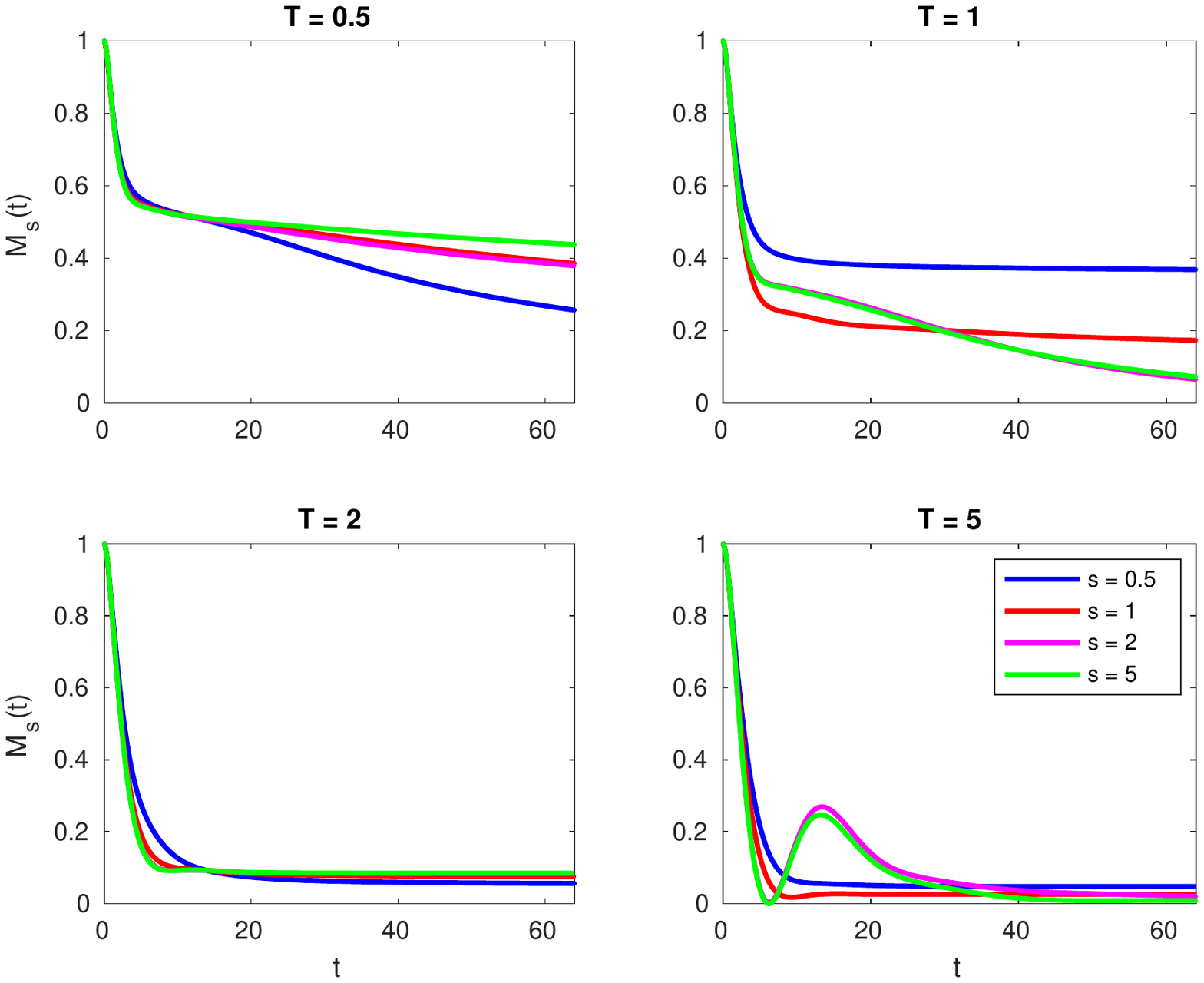}
\caption{Evolution with time of scaled mix-norm $M_s(t)$, as defined in (\ref{eq:msdef}) for flows initially seeded with perturbations that minimize mix-norms (for a variety of the values of index $s$) over different target times $T$ for flows with $Re=Pe=500$.  }\label{fig:mix500}
\end{figure}
We now compare and contrast the behaviour of the flows with $Pe=500$ with the flows associated with  $Pe=50$. The evolution of the various mix-norms $M_s(t)$ for flows with $Pe=500$ are shown in figure \ref{fig:mix500}. (The equivalent time evolution of the variance for these flows is shown in figure \ref{fig:var500}.)   
For the $T=0.5$ fields we observe that initially all $M_s(t)$ follow the same decay for early times $t \lesssim 20$. After this initial decay, there is some separation, with $M_{0.5} (t)$  decaying the quickest. 
The solutions for $T = 1$ do not follow this behaviour,
with $M_{2}(t)$ and $M_{5}(t)$ eventually decaying the quickest. Interestingly, the various optimal perturbations
with target time $T=2$ behave in a very similar fashion, all
leading to very small values at relatively early times. 
Finally,  the qualitative picture is different again for the various perturbations for the target time $T=5$. As seen in the low P\'eclet number case, demixing (in that $M_s(t)$ is non-monotonic) occurs for the larger indices $s=2$ and $5$, clearly associated with the `bump' observed in the fourth panel of figure \ref{fig:var500}.

As in the case of $Pe=50$, variation of the control parameters produces qualitatively different mixing for flows with $Pe=500$, with again a tendency for 
optimal perturbations with short target times being associated with smaller scales and more rapid dissipation of perturbation kinetic energy.  
\begin{figure}[!h]
    \centering
    \includegraphics[width=0.9\textwidth]{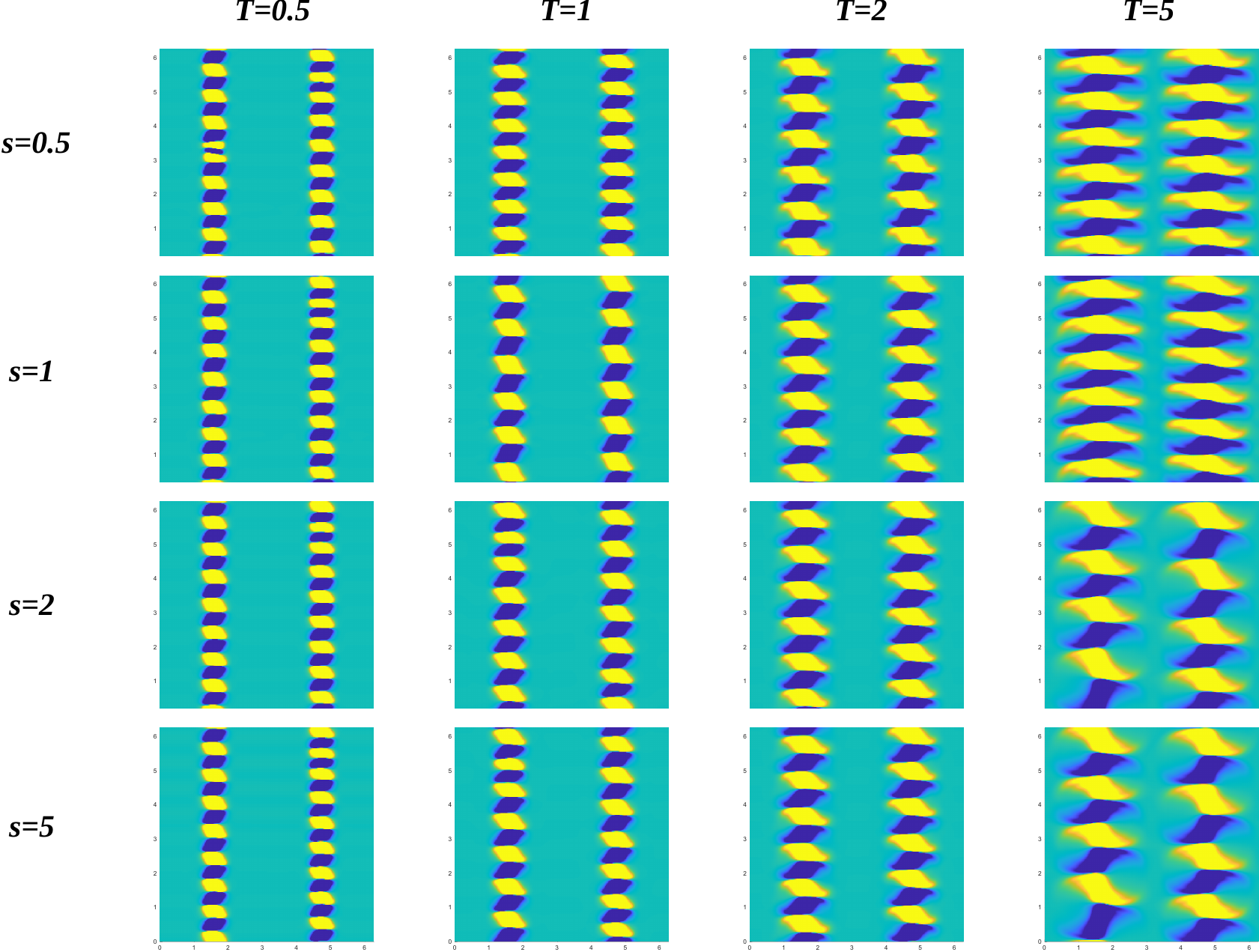}
    \hspace*{1.2cm}
    \includegraphics[width=0.6\textwidth]{colorbar-crop.pdf}
\caption{Vorticity distribution of the optimal 
perturbations identified for mix-norm minimization for flows with $Pe=500$ with various combinations of index $s$ (rows) and target time $T$ (columns).}\label{fig:vort500}
\end{figure}

To investigate further the differences and similarities between flows with $Pe=50$ and $Pe=500$, we plot in figure \ref{fig:vort500} the initial vorticity of the various optimal perturbations for $Pe=500$. Similarly to the optimal initial perturbations for $Pe=50$, increasing target time decreases the number of vortices and thus increases the length scale. With the exception of the $T = 5$ case, changing index does not appear to have as large an effect on the initial structure, in contrast to the optimal perturbations for flows with $Pe=50$ (shown in figure \ref{fig:vort50}).  

An apparent difference between the perturbations
for lower and higher $Pe$ is the shape of the vortices. For initial perturbations in flows with $Pe=50$, the vortices are (close to) circular counter rotating vortices aligned along the interface between regions of low and high passive scalar. However, this is not the case for the perturbations in flows with  $Pe=500$, where the vortices take more of a \textit{diamond} quadilateral shape, with finer-scale structure being apparent, with two   different types of  structure for the perturbations associated with shorter and longer target times. 

For the perturbations associated with shorter target times,  the vortices take on a quadrilateral shape with sharp corners, while for longer target times, two of the quadrilateral `corners' become somewhat elongated. 
\begin{figure}[!h]
    \centering
    \includegraphics[width=0.9\textwidth]{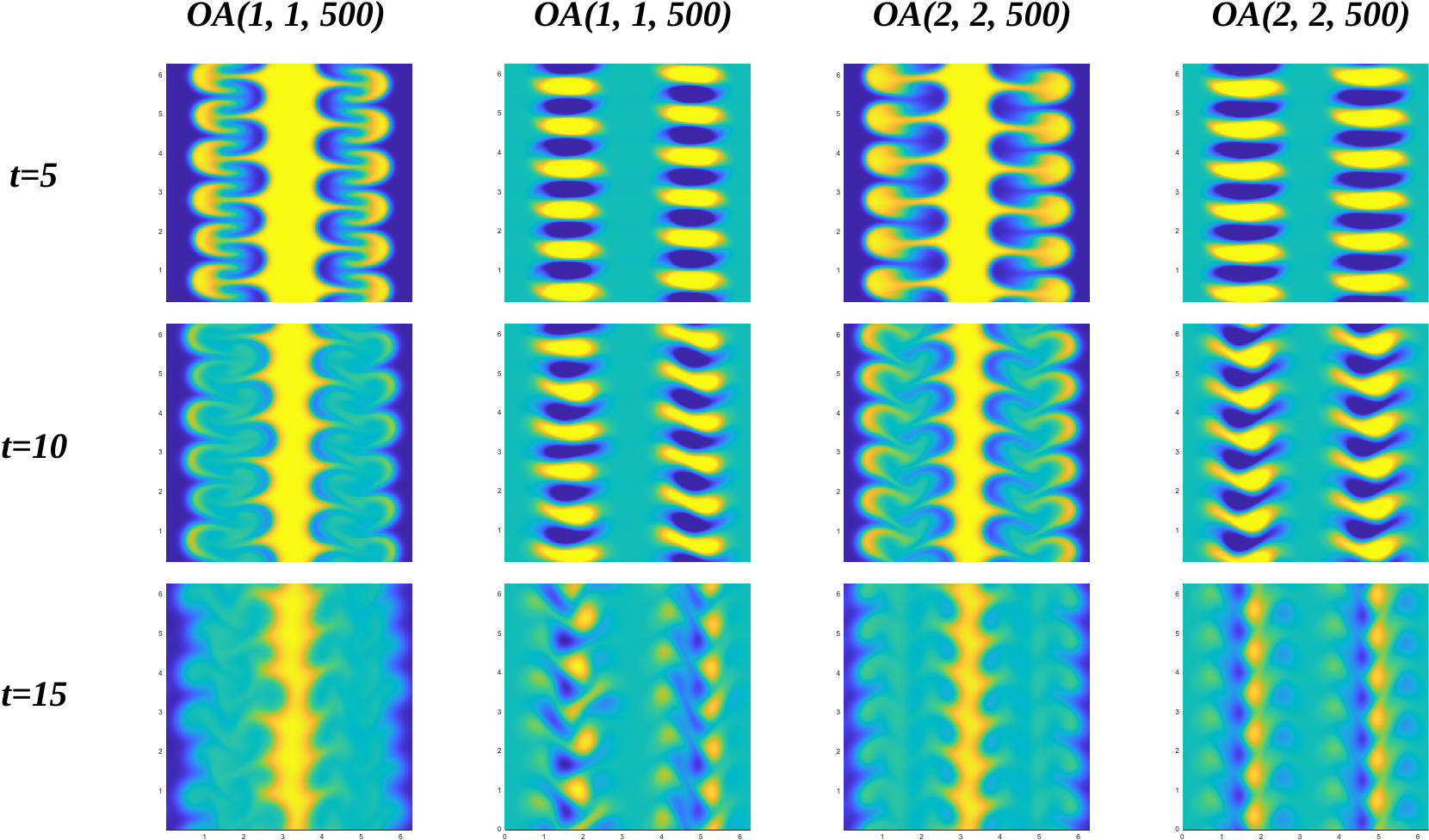}
    \hspace*{1.2cm}
    \includegraphics[width=0.6\textwidth]{colorbar-crop.pdf}
\caption{Snapshots of the evolution of the passive scalar (left) and vorticity (right) for the fields $OA(1,1,500)$ on the first two columns and $OA(2,2,500)$ on the second two columns. Snapshot times are $t=5,10,15$.}\label{fig:t2500}
\end{figure}

To investigate the difference between these two structures, we plot in figure \ref{fig:t2500} the evolution of the vorticity and passive scalar for optimal perturbations $OA(1,1,500)$ (with the `pure' quadrilateral initial vortices) and $OA(2,2,500)$ (with the `cornered' initial vortices). (Once again, animations are available as supplementary materials.)
For both these  higher P\'eclet number flows, the vortices stretch and fold the interfaces of the passive scalar  effectively by $t = 5$. The vortices then deform to further increase the stretching and folding of the interface, acting with diffusion to mix the fluid. The difference between the `pure' and `cornered' quadilateral initial 
vortices becomes more apparent in the range $10 < t < 20$,
with a clearer `V' developing in the flow
associated with the $OA(2,2,500)$ initial perturbations,
which remain more organised than those
associated with the $OA(1,1,500)$ initial perturbations,
and thus more able to mix the initial vertical striping of {scalar. However}, for both flows, it is apparent that the scalar field 
is imperfectly mixed, with the initial vortices still remaining too small to disrupt and homogenize entirely the initial vertical striping of the scalar field. 

\begin{figure}[!h]
    \centering
    \includegraphics[width=0.9\textwidth]{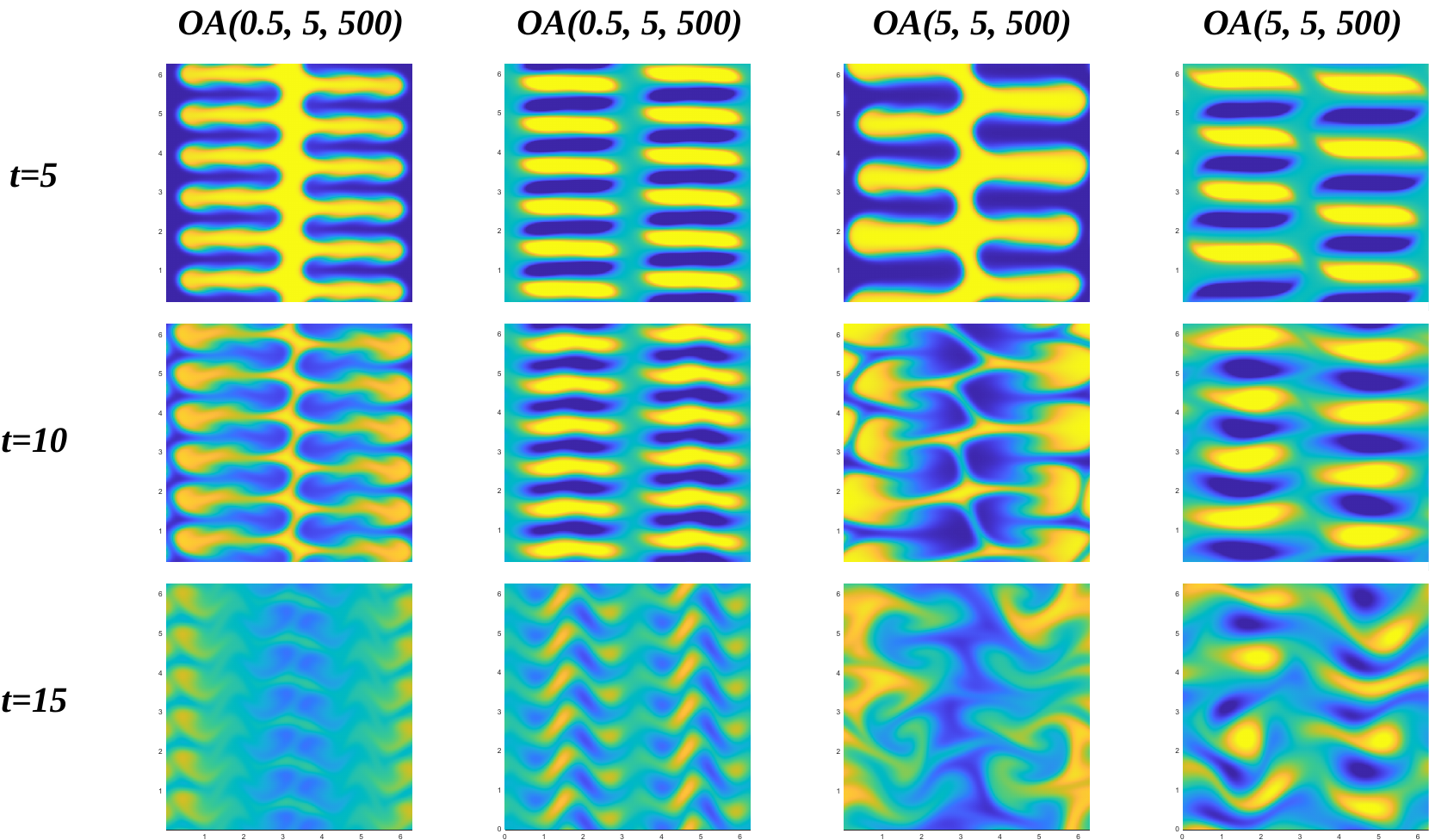}
    \hspace*{1.2cm}
    \includegraphics[width=0.6\textwidth]{colorbar-crop.pdf}
    \caption{Snapshots of the evolution of the passive scalar (left)  and vorticity (right) for the optimal perturbations $OA(0.5,5,500)$ (shown in the  two left-most columns) and $OA(5,5,500)$ (shown in the two right-most columns). Snapshot times are $t=5,10,15$.}
    \label{fig:t5500}
\end{figure}

In flows associated with the longest target time considered, i.e. $T=5$ with $Pe=500$, the dynamical significance of the `cornered' initial vortices becomes apparent. In figure \ref{fig:t5500}, time evolution of the flows
associated with $OA(0.5,5,500)$ (left-most columns) and
$OA(5,5,500)$ (right-most columns) are shown. 
As can be seen from figure \ref{fig:mix500}, the $OA(5,5,500)$ exhibits 
`demixing' behaviour, which can now be understood in terms of the observed physical flow evolution. 
For both the flows shown in figure \ref{fig:t5500}, the cornered vortices become sufficiently horizontally elongated
to perturb the entirety of the scalar field. For the flow evolving from the optimal initial perturbation $OA(0.5,5,500)$, the vortices are both sufficiently small and sufficiently regular to lead to a smooth, organised, perturbation and diffusion of the scalar field, leading to homogenization at relatively early time. On the other hand, the vortices associated with the $OA(5,5,500)$ are larger, and so actually remain more coherent at $t=15$. As they remain (more) coherent, they  advect the scalar field too strongly, leading to an inversion in the vertical striping, leading to a negative scalar field stripe (shown in blue) in the middle of the flow domain at intermediate time,
an even stronger demixing effect than seen before in figure
\ref{fig:diag50} for the $Pe=50$ flows with high index $s=5$.
This organised inversion of the scalar field distribution manifests itself in the non-monotonic variation in the mix-norm $M_5(t)$ as shown in figure \ref{fig:mix500}, and demonstrates
that the `best' choice of index for such optimization 
problems is in general likely to be $Pe$-dependent, particularly if the mixing problem of interest has a finite
time horizon of interest. 

\section{Conclusions}
We have used the DAL method  to identify optimal initial perturbations for passive scalar mixing in a 2D toroidal geometry for different choices of the key  parameters: $s$ the index of the mix-norm, target time $T$, and P\'eclet number of the flow $Pe$. We have demonstrated that   mix-norms (with various indices)  can indeed be used as a proxy for variance-based strategies,
which both converge relatively rapidly and robustly identify 
flow evolutions that minimise variance over times significantly longer than the chosen target time $T$. 
We have also investigated the qualitative change in mixing dynamics at both low and high P\'eclet number, demonstrating
substantial qualitative variability both in the optimal initial perturbations and the subsequent flow evolution. 
Specifically, for flows with  $Pe=50$, highly symmetrical arrangement  of initial vortices  leads to optimal mixing.

However, for flows  with $Pe=500$, there
is evidence of a trade-off, in that larger, highly organized vortices can actually lead to `{demixing}', with too vigorous advection not allowing diffusive processes
to homogenize the scalar distribution, at least at intermediate times. Since higher values of the mix-norm index strongly deprecate small scales, this behaviour
strongly suggests that the mix-norm
is most useful in such mixing optimization problems
when the index used is not too large, i.e. choosing $s\sim 1-2$ seems the most practical choice. It would clearly be of interest if that could be established rigorously.  

We therefore conjecture that the mix-norm may be used to identify organized initial perturbations that lead to a flow evolution that is optimal for mixing (for a given initial energy cost), and highlighted, particularly for the flows with $Pe=500$, the importance of initial vortical perturbations with fine-scale structure, which we referred to as `corners', apparent in figure \ref{fig:vort500} for higher target times. However, initial energy density and Schmidt number were kept constant at $e_0=0.03$ and $Sc=1$ respectively. We suggest that further studies should vary these parameters as well to get a more comprehensive picture of the structure of perturbations that actually lead to optimal mixing. We also suggest that varying the index in different geometries and applying an external force to compare with existing studies.

We conclude with the observation that despite having different indices, different mix-norms lead to very similar temporal evolution of variance (and hence mixing properties). This appears to be evidence in support of  a hypothesis from \cite{21} that there exist a range of indices which decay at very similar rates, although of course further detailed investigation is necessary. 

\vskip6pt

\enlargethispage{20pt}

\textbf{Data Accessibility:} Source code, numerical data and animations of the flows shown in figures 5, 6, 9 and {10} have been provided as electronic supplementary material.

\textbf{Contributions:} All authors contributed equally to the present article.

\textbf{Competing Interests} We declare we have no competing interests.

\textbf{Acknowledgements} This work was done as part of a PhD project made possible by the National University of Ireland's Travelling Studentship in the Sciences and their support is gratefully acknowledged by the authors. We dedicate this paper to Charlie Doering, who first suggested to C. P. C. the potential for mix-norm optimization to be computationally efficient. We would like to thank Luk\'{a}\v{s} Vermach whose work inspired this project. We would also like to thank Florence Marcotte for some helpful comments on the optimization algorithm.


\end{document}